\documentclass[floats,twocolumn,aps,epsf,prl]{revtex4}
\usepackage{graphicx}

\def \be{\begin{displaymath}}
\def \ee{\end{displaymath}}              

\def \ben{ \begin{equation} }
\def \een{ \end{equation}   }            

\def \bea{\begin{eqnarray*}}             
\def \eea{\end{eqnarray*}}

\def \bean{\begin{eqnarray}}             
\def \eean{\end{eqnarray}}

\def \invb#1 { \frac{1}{#1} }

\begin{document}

\title{Proposed experiments to probe the non-abelian $\nu=5/2$ quantum Hall state}

\author{Ady Stern$^{(1)}$ and Bertrand I. Halperin$^{(2)}$}

\address{$^{(1)}$ Department of Condensed Matter Physics, The Weizmann Institute of Science,
Rehovot 76100, Israel\\ $^{(2)}$ Physics Department, Harvard University, Cambridge, Ma 02138}



\begin{abstract}
  
We propose several experiments to test the non-abelian nature of quasi-particles in the fractional quantum Hall state of $\nu=5/2$. One set of experiments studies interference contribution to back-scattering of current, and is a simplified version of an experiment suggested recently \cite{DFN}. Another set looks at thermodynamic properties of a closed system. Both experiments are only weakly sensitive to disorder-induced distribution of localized quasi-particles.

\end{abstract}

\pacs{PACS}

\maketitle

\parindent=0.8cm


Non-abelian quantum Hall states have been the focus of much theoretical interest since their 
proposal by Moore and Read \cite{Moore,Morf,Read-review}.
This interest has been recently revived for several reasons. First, improved experimental capabilities allow for better inspection of the quantum Hall states in the range of Landau level filling fractions of $2<\nu<4$, where at least some of the observed states may be non-abelian\cite{Pan}. Second, non-abelian quantum Hall states are believed to be abundant in rotated Bose-Einstein condenstaes, at high (presently unattainable) angular rotation velocity\cite{Cooper}\cite{ReadRezayi}. And third, non-abelian quantum Hall states are prime candidates for the realization of a topological quantum computer\cite{Kitaev}. 

Experimental study of the non-abelian nature of quantum Hall states has so far been lacking, both because of difficulty in reaching the experimental conditions needed for this study (particularly the quality of the two dimensional electron gas), and because of the lack of proposals for realizable experimental tests. The quality of samples was impressively improved in recent years, and the need for proposals for experiments becomes ever more burning. Important steps in that direction were carried out by Fradkin et al.\cite{Fradkin}, who considered an interferometer for non-abelian quasi-particles and pointed out its general relation to Jones polynomials, and by Das Sarma et al., who proposed an interference experiment whose results test the non-abelian nature of excitations in the $\nu=5/2$ state\cite{DFN}. 

In this work we propose several simplified versions for such an experiment, and examine the conditions under which it may indeed be a test for the non-abelian nature of the $\nu=5/2$ state. We focus first on a Hall bar where two quantum point contacts introduce weak back-scattering of current, with amplitudes of $t_{\rm L}$ and $t_{\rm R}$ respectively (see Fig. [1]). In the simplest case, where the bulk is in an integer quantum Hall state, one expects the back-scattered current to be proportional to $|t_{\rm L}\ +\ e^{i2\pi\Omega}\ t_{\rm R}|^2$ where the relative phase $\Omega$ is the number of flux quanta enclosed in the island defined by the two quantum point contacts and the two edges connecting them. This phase can then be varied either by a variation of the magnetic field or by a variation of the area of the island, e.g., by means of a side gate. When the back-scattering is measured as a function of one of these two parameters, an interference pattern is obtained, with a period corresponding to one flux quantum. In the case of the fractional quantized Hall state at $\nu=1/3$, when the interference pattern is measured by varying the size of the island, its period corresponds to a change of three in the number of flux quanta enclosed by the island \cite{Sondhietal}, reflecting the fact that the quasiparticles which tunnel across the point contacts carry a fractional charge of $e/3$.

\begin{figure}[h]
\begin{center}
\includegraphics[width=8.0cm]{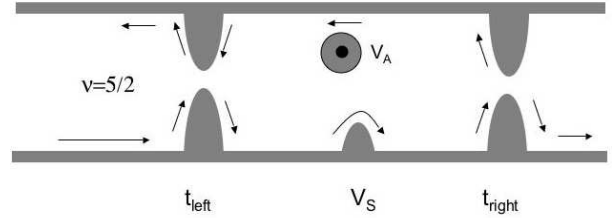}
\end{center}
\caption{Experimental set-up for measuring the interference contribution to the back-scattered current. Current flows along the lower edge, heading rightwards, and is backscattered by two quantum point contacts. The "island" is defined by the two quantum point contacts and the two edges. The anti-dot at the center of the island is coupled to an air-bridge gate that controls the number of $e/4$-charged quasi-particles that it localizes. A voltage $V_S$ applied to a side gate varies the size of the "island".}\label{Fig1}
\end{figure}


We analyze this experiment for the $\nu=5/2$ case, and find that it reflects the special character of this state in two ways. As in the abelian FQHE states, the period of the oscillations, when measured by varying the area of the island, reflects the $e/4$ charge of the elementary excitations. More interestingly, however, we consider the effect of $e/4$--charged quasi-particles localized statically within the island. We denote by $n_{\rm is}$ the number of these quasi-partciles. We find that as a consequence of the non-abelian character of the quasi-particles, the oscillations are suppressed when $n_{\rm is}$ is odd, and are revived when this number is even. 

Following this analysis, we consider the limit of strong back--scattering, where the island becomes a Coulomb--blockaded quantum dot. Measurements of the conductance through the dot allow in this limit an extraction of its addition spectrum. We find similar sensitivity of the addition spectrum to the parity of $n_{\rm is}$. 

In order to establish our results and study their consequences, we start by reviewing the basic theory of the non-abelian $\nu=5/2$ state\cite{ReadGreen}. The Moore-Read non-abelian $\nu=5/2$ quantum Hall state 
may be regarded as a $p$-wave super-conductor of composite fermions. Precisely at $\nu=5/2$, at zero temperature $T=0$, this super-conductor is free of topological defects. At a filling factor $\nu=\frac{5}{2}+\epsilon$, with $|\epsilon|\ll 1$, the super-conductor is pierced by well-separated vortices. Then, each vortex $i$ carries a zero-energy mode $\gamma_i$ localized to its core. These modes may be written as Majorana fermions
\begin{equation}
\gamma_i=\int d{\bf r}\left [g_i({\bf r})\psi({\bf r})+g_i^*({\bf r})\psi^\dagger({\bf r})\right]
\end{equation}
where $\psi(\bf r)$ annihilates a composite fermion at point $\bf r$. 
The function $g_i$ is  localized at the $i$'th vortex core but has a phase that depends on the position of all other vortices. 
The Majorana operators satisfy $\{\gamma_i,\gamma_j\}=2\delta_{ij}$. 
As a consequence, when the vortices are pinned to their position, the ground state becomes degenerate. For $2n$ quasi-particles located at $\{R_1...R_{2n}\}$, the ground state subspace is of dimension $2^{n}$, and it is spanned by the wave functions $\Psi_k\{R_1...R_{2n}\} $, in which the vortex positions are parameters, and the integer index $1\le k\le 2^n$. 
A braiding of the vortex positions $R_j$'s is accompanied by a unitary transformation acting within this subspace. These transformations are conveniently expressed in terms of the Majorana operators $\gamma_i$. In particular, when a vortex $i$ encircles vortex $j$, the unitary transformation is, up to a phase factor, $\gamma_i\gamma_j$ \cite{NayakWilczek,IvanovStern}. 

When a quasi-particle comes from $x=-\infty$ along the right moving edge, gets back-scattered by one of the two point contacts, and moves back to $x=-\infty$ along the left-moving edge, it may end up encircling along its way the $n_{\rm is}$ quasi-particles localized within the bulk. In the presence of such localized quasi-particles, when the quasi-particle moving along the edge comes back to $x=-\infty$, it leaves the system in a ground state different from the one it started at. We denote the initial ground state as $|{\rm initial}\rangle$, the unitary transformation applied on that state by the partial wave scattered by the right point contact as $U_{R}$, and the unitary transformation applied by the partial wave scattered at the left point contact as $U_L$. With this notation, the $\Omega$-dependent part of the back-scattered current becomes, to lowest order in the back-scattering amplitudes, 
\begin{equation}
\label{backscat}
2{\rm Re}\left[ t_{\rm L}^*t_{\rm R}e^{2\pi i\Omega}\langle {\rm initial}|U_L^{-1}U_R|{\rm initial }\rangle \right]
\end{equation}
The definition of $\Omega$ involves an arbitrary additive constant, since $t_{\rm L},t_{\rm R}$ are complex. The variation of $\Omega$ with the area $A$ of the island satisfies,
\begin{equation}
\label{Omega}
\frac{\partial\Omega}{\partial A} = B/4\Phi_0 =  n_0/2,
\end{equation}
where $B$ is the magnetic field, $\Phi_0$ is the flux quantum, and $n_0$ is the density of electrons in the partly filled Landau level, as  expected for quasiparticles with charge $e/4$.
Denoting the Majorana mode of the interfering quasi-particle by $\gamma_a$, and those of the $n_{\rm is}$ quasi-particles localized at the island between the point contacts by $\gamma_i$, with $1\le i\le n_{\rm is}$, we find that up to a possible phase which is independent of $\Omega$ and of the ground state at which the system is, 
\begin{equation}
U_a\equiv U_l^{-1}U_r=\gamma_a^{n_{\rm is}}\Gamma
\label{uliur}
\end{equation}
with $\Gamma\equiv \prod_{i=1}^{n_{\rm is}}\gamma_i$. For our future discussion it is useful to note that the operator $U_a^2=(-1)^{[3n_{\rm is}/2]}$, with $[3n_{\rm is}/2]$ denoting the integer part of $3n_{\rm is}/2$. Thus, $U_a$ has two eigenvalues that differ from one another by a minus sign (either $\pm 1$ or $\pm i$).

It is in the expression (\ref{uliur}) that the parity of $n_{\rm is}$ becomes crucial. Since for any even $n_{\rm is}$ we have $\gamma_a^{n_{\rm is}}=1$, the product (\ref{uliur}) of the two unitary transformations is independent of the Majorana mode of the incoming particle $\gamma_a$, and {\it is the same for all incoming particles}.  

The two eigenvalues of $\Gamma$ correspond to two interference patterns that are mutually shifted by $180^0$. If the initial ground state $|{\rm initial }\rangle $ is an eigenstate of $\Gamma$ then the phase of the interference pattern is determined by the corresponding eigenvalue. Furthermore, even if $|{\rm initial }\rangle $ is not an eigenstate of $\Gamma$ an interference pattern will be observed \cite{Bais}. In that case the electronic current driven from $x=-\infty$ and being backscattered from the two point contacts acts as a measuring device of $\Gamma$, as when enough quasi-particles flow through the system to ascertain the backscattering probability the eigenvalue of $\Gamma$ may be extracted from that probability. Thus, a measurement of the two terminal conductance of the system turns the superposition of different eigenvalues of $\Gamma$ into a mixed state, and in effect collapses the system to one of the eigenvalues of $\Gamma$. 


The physical distinction between the two subspaces that correspond to the two eigenvalues of $\Gamma$ becomes clearer if one considers the limit of strong backscattering at the constrictions of Fig [1]. where the island becomes a closed system. Then, the subspace spanned by $\gamma_i$ ($i=1..{n_{\rm is}}$) is split to two subspaces of equal dimension ($2^{\frac{n_{\rm is}}{2}-1}$) that correspond to a different parity of the total number of electrons in the closed island. The eigenvalues of the operator $\Gamma$ distinguish between these two subspaces. In a closed system, this eigenvalue cannot be changed by operations that involve braiding between the $n_{\rm is}$ localized quasi-particles. Similarly, in the open system, one needs a quasi-particle exterior to the $n_{\rm is}$ localized ones to tunnel between the edges in order to change that eigenvalue (see Ref. \cite{DFN}, as well as the discussion towards the end of this paper). 

The effect of the localized quasi-particles on the interference is very different when $n_{\rm is}$ is odd. Then, the unitary transformation $U_a$, Eq. (\ref{uliur}), includes the Majorana operator of the interfering quasi-particle. For two different incoming quasi-particles $a,b$ the operators $U_a,U_b$ do not commute. Rather, $[U_a,U_b]=(-1)^{[3n_{\rm is}/2]}\gamma_a\gamma_b$. Thus, for each incoming quasi-particle the interference term is multiplied by a different factor, and over all, the interference is dephased, and {\it the back-scattered current} (\ref{backscat}) {\it becomes independent of $\Omega$}. 

As shown above, the interference pattern that is to be observed in the set-up of Fig. (1) depends crucially on the parity of $n_{\rm is}$. For the observation of such a dependence we need to increment $n_{\rm is}$ in a controlled fashion. To that end, we note that quasi-particles are introduced into the system by local deviations of the filling factor from $\nu=5/2$. We consider three experimental knobs for the mapping of the interference pattern. The first is the side gate in Fig. (1). When the voltage on that side gate, $V_S$, is varied, the size of the island varies, but (ideally) the electron density and the filling factor are unchanged in the interior.  Then $\Omega$ is varied, but no new quasi-particles are introduced. The second knob is the anti-dot near one edge of the island (see Fig. (1)). We assume this anti-dot to be small enough such that its charging energy is larger than the temperature and therefore its charge is quantized in units of $e/4$ by the Coulomb blockade. Furthermore, we assume this charge to be variable by means of an air-bridged gate that couples to the anti-dot. When the voltage on that gate, $V_A$, is varied, the number of $e/4$ quasi-particles charging the anti-dot is varied, and thus so is also $n_{\rm is}$. The matrix element for quasiparticles to tunnel between the antidot and the edge of the island should be large enough so that tunneling can occur when the gate voltage is swept through the resonant condition but negligible when the antidot is off resonance. The third knob is the magnetic field.

There are two experimental procedures to test the effect of the parity of $n_{\rm is}$. In the first procedure the magnetic field is kept fixed, and the back-scattered current is measured as a function of the size of the island and the voltage $V_A$. We expect oscillations of the back-scattering current as a function of $V_S$, and we expect the amplitude of these oscillations to vary discontinuously with $V_A$, changing periodically between zero and $O(t_{\rm L}^*t_{\rm R})$ as $n_{\rm is}$ is varied with $V_A$.

In the second procedure we turn off the anti-dot and vary $n_{\rm is}$ by varying the magnetic field $B$. If the density is kept fixed, the variation of $B$ changes the filling factor uniformly within the island. For small deviations from $\nu=5/2$, a set of localized quasi-particles will be introduced into the island, and $n_{\rm is}$ will vary with $B$. The positions and the precise values of $B$ at which these quasi-particles will enter the island depend on the precise shape of the island and the disorder potential it encompasses. On average, a change in the magnetic field by one tenth of a flux quantum introduces one $e/4$-charged quasi-particle into the island, but fluctuations from that rate are to be expected. In any case, the back-scattered current should again oscillate with $V_S$, and the amplitude of these oscillations should be turned on/off with the introduction of quasi-particles by the variation of $B$. 

So far we have assumed that the parity of $n_{\rm is}$ is time-independent throughout the experiment. For that assumption to be valid, the charge on the island should have typical fluctuations much smaller than $e/4$, or, if this condition is not realized, fluctuations whose characteristic time scale is much longer than that of the experiment. Assuming that the conductance of the island to the outside bulk, $G$, is frequency independent, and confining our attention to frequencies $\omega\ll T/\hbar$, the charge fluctuations on the island satisfy 
\begin{equation}
\langle Q(t=0)Q(t)\rangle =2CT\exp{-t/\tau}
\label{chargeflucts}
\end{equation}
where $C$ is the capacitance of the island and $\tau=C/G$ is the relaxation time for the charge fluctuations. For the thermal fluctuations of the charge to be much smaller than $e/4$, then, the capacitance of the island should satisfy $C\ll \frac{e^2}{32 T}$. The determination of the relevant capacitance is rather subtle, however, since the bulk of the island, where the $n_{\rm is}$ quasi-particles are located, is electrostatically coupled to the edges, to the bulk outside of the island, and to various external gates. 

Analogous phenomena may be observed in closed systems at $\nu=5/2$. We consider a closed island with $n_{\rm is}$ pinned quasi-particles in its bulk, and study the way the energy of the island varies when its area is varied by the application of a voltage $V_S$ to a side gate. For simplicity, we first disregard the two filled Landau levels, and incorporate their effect later. We assume a very weak coupling of the island to an electron reservoir, such that as the area is varied, the number of electrons in the island varies as well, but for any fixed $V_S$, this number is fixed to an integer. 

Since the $\nu=5/2$ state is a super-conductor of composite fermions, and the number of composite fermions is identical to the number of electrons, one may naively expect even-odd oscillations of the energy as a function of the number of electrons, reflecting the difference between a fully paired ground state of a super-conducting island and one in which one electron is unpaired\cite{Tuominen}\cite{Devoret}: When the number of electrons in such an island is increased from an even number to an odd one, the extra electron occupies the lowest BCS quasi-particle mode, whose energy is gapped from the ground state. A further increase back to an even number of electrons does not require the creation of BCS quasi-particles, since all electrons become paired in the condensate. 

The $p$--wave super-conductor of composite fermions that we discuss here is rather unconventional in having a sub-gap excitation branch near the edge \cite{ReadGreen}. When the area of the island is increased and electrons are added, the unpaired electrons, if any, occupy the lowest state of that branch. And it is in the energy of that state, which we denote by $\delta_0$, that the parity of $n_{\rm is}$ has an effect: When $n_{\rm is}$ is odd, $\delta_0=0$, and the dependence of the energy on the number of electrons does not show any even-odd effect. In contrast, when $n_{\rm is}$ is even, $\delta_0$ is small (inversely proportional to the perimeter of the island), {\it but non-zero}\cite{ReadGreen}. In this case, as the number of electrons in the island is increased the energy cost for adding an electron depends on whether the added electron is paired or un-paired. For a closed system, then, the even-odd effect in the energy cost associated with changing the electron number by one is turned off when $n_{\rm is}$ is odd and turned on when $n_{\rm is}$ is even. 

A practical way of measuring this even-odd effect may use the system in Fig. (1) in the limit of strong back-scattering, at which a quantum dot of $\nu=5/2$ is formed between the point contacts, and the side gate is used to vary the dot's area. If the two terminal conductance of the dot is measured as a function of the gate voltage $V_S$, the well-known series of conductance peaks is to be expected, associated with values where the number of electrons on the dot changes by one. The voltage separation between these peaks measures the energy cost involved in adding an extra electron, and is the quantity that should reflect the parity of $n_{\rm is}$.  In the case where $n_{\rm is}$ is odd, the average spacing between peaks corresponds to an area change $\delta A = 1/n_0$. In the case where it is even, there will be even-odd fluctuations about this average, so that the true period becomes $2/n_0$. The parity of $n_{\rm is}$, in turn, may be varied by a variation of the magnetic field. Again, on average $n_{\rm is}$ is varied by one when the flux through the dot is varied by $1/10$ of a flux quantum. 

For a closed island, a change in $V_S$ affects also the occupation of the two filled Landau levels, and introduces additional Coulomb blockade peaks associated with this occupation. The periodicity of these peaks corresponds to $\delta A=1/n_0$, and therefore does not eliminate the distinction between odd and even $n_{\rm is}$.

The two limits we discuss, of weak and strong back--scattering, may be compared through a Fourier decomposition of the conductance as a function of the area of the island. We write 
$G = \sum_m g_m e^{2 \pi i m \Omega}$, where $\Omega$ is defined by (\ref{Omega}). Terms with $m$ odd should be absent when $n_{island }$  is odd and present when $n_{\rm is}$ is  even. In the limit of weak backscattering, successive terms get smaller by the small factor of $t_{\rm L}^*t_{\rm R}$, while in the limit of strong back--scattering no such small parameter exists.  

Since for a closed system the number of electrons is quantized to an integer, thermal fluctuations of the charge on the island must be much smaller than the electron charge, and thus the capacitance of the dot should satisfy $C\ll \frac{e^2}{2 T}$. Furthermore, the temperature should be smaller than $\delta_0$.

The experiments suggested so far, examining the way that the interference contribution to back-scattering and the energy cost for adding an electron are affected by the parity of $n_{\rm is}$, are simpler than the experiment suggested by Das Sarma {\it et al.}\cite{DFN}. The simplicity is a consequence of the less ambitious goal we address. In the Das Sarma {\it et al.} proposal $n_{\rm is}=2$, and the set-up is devised such that at the crucial part of the experiment the eigenvalue of $\Gamma$ is changed and the interference pattern is shifted  by $180^0$. In practice, it is probably very difficult to tune $n_{\rm is}$ precisely to two, due to the unavoidable abundance of localized quasi-particles resulting from density non-uniformities. As our analysis above shows, when $n_{\rm is}$ is odd, no interference is to be observed, and no phase shift may be induced. We further examine the case when $n_{\rm is}$ is an even number different from two. In that case a measurement of the back-scattered current as a function of  the side gate voltage $V_S$ collapses the system into a ground state with a particular eigenvalue of $\Gamma$, one of the two possible eigenvalues. In order to change that eigenvalue we need to apply a unitary transformation $\tilde\Gamma$ that does not commute with $\Gamma$. It is easy to see that if the even-numbered $n_{\rm is}$ localized quasi-particles are separated into two odd-numbered groups of quasi-particles, and a quasi-particle from the edge encircles the quasi-particles of one of these two groups (say group number $1$), the resulting unitary transformation $\gamma_a\prod_i^{(1)}\gamma_i$ (where $\prod_i^{(1)}\gamma_i$ indicates a product over all Majorana operators of the quasi-particles of group number $1$) changes the eigenvalue of $\Gamma$. As suggested by Das Sarma et al., this transformation may be applied by a single quasi-particle tunneling between the edges through another quantum point contact, situated between the left and right ones. This point contact divides the island into two parts. The present analysis reveals, then, that the Das Sarma et al. procedure would indeed shift the interference pattern by $180^0$ {\it only if each of these two parts includes an odd number of quasi-particles}. 

To summarize, in this paper we propose several experiments that probe the non-abelian character of the $\nu=5/2$ quantum Hall state, both through transport measurements in an open system and through thermodynamic measurements in a closed system\cite{BSK}. 

We thank A. Kitaev and CM Marcus for instructive discussions. We acknowledge support from the US-Israel Binational Science Foundation, the Minerva foundation (AS), and NSF grant DMR 02-33773 (BIH).

\end{document}